%% file: text.tex
\documentclass{PoS}
\usepackage{booktabs,graphicx,dcolumn}

\newcommand{\mathe}{\mathrm{e}}

\newcommand{\tmtextbf}[1]{{\bfseries{#1}}}
\newcommand{\tmtextit}[1]{{\itshape{#1}}}
\newcommand{\tmtexttt}[1]{{\ttfamily{#1}}}

\title{Triviality of $\phi^4_4$ in the broken phase revisited}

\ShortTitle{Triviality of $\phi^4_4$ in the broken phase revisited}

\author{\speaker{Tomasz Korzec and Ulli Wolff}%
\\
        Humboldt Universitat zu Berlin, Institut fur Physik, Newtonstrasse 15, D-12489 Berlin, Germany\\
        E-mail: \email{korzec@physik.hu-berlin.de, uwolff@physik.hu-berlin.de}}

\abstract{We define a finite size renormalization scheme for
$\phi^4$ theory which in the thermodynamic limit
reduces to the standard scheme used in the
broken phase. We use it to re-investigate
the question of triviality for the four dimensional
infinite bare coupling (Ising) limit. The relevant
observables all rely on two-point functions and
are very suitable for a precise estimation 
with the worm algorithm. This contribution updates
an earlier publication by analysing a much larger dataset.

\begin{flushright} HU-EP-15/04\\SFB/CPP-14-117 \end{flushright}
}

\FullConference{The 32nd International Symposium on Lattice Field Theory\\
                 23-28 June, 2014\\
                 Columbia University New York, NY}

\begin{document}
\section{Summary of theoretical background}
This brief article gives an update on \cite{Siefert:2014ela} in the sense that
we here analyze a very much enlarged new data set. This is found in the next section while
we here summarize the theory for the reader's convenience\footnote{
A more detailed elementary introduction can be found in \cite{UWscholped:2014}.
}.

We consider the single component Z(2) symmetric scalar field theory on a torus of length $L$
embedded in four dimensional Euclidean space. We employ the simplest hypercubic
lattice discretization with $L/a$ sites in each direction and the
standard nearest neighbor lattice action
\begin{equation}
  S = \sum_x \left[ \varphi \left( x \right)^2 + \lambda \left( \varphi \left(
  x \right)^2 - 1 \right)^2 \right] - 2 \kappa \sum_{x \mu} \varphi \left( x
  \right) \varphi \left( x + \hat{\mu} \right).
\end{equation}

The standard picture for this quantum field theory is \cite{Luscher:1987ay,Luscher:1987ek} 
that there is a critical
line $\kappa_c(\lambda)$ where the model possesses a continuum limit. This limit
in $\kappa$ may be approached from below to reach the symmetric massive continuum
theory or (for $\lambda >0$) from above to define the broken-symmetry massive model. The latter is of
particular theorectical interest here due to some similarity to the Higgs field in the 
standard model.

As a standard way to renormalize the
infinite volume
theory in the broken phase we may match
the Fourier transform $G(p)$ of
the two point correlation function
\begin{equation}
G(p)=\sum_x \mathe^{-ipx} \langle \varphi(x) \varphi(0) \rangle
\end{equation}
to the asymptotic form
\begin{equation}\label{RNsmallp}
G(p) = Z\left\{ v^2 \delta^4(p) + \frac1{\hat{p}^2+m^2+\mathrm{O}(\hat{p}^4)} \right\}
\quad (p \to 0)
\end{equation}
in the limit of vanishing $p$.
Here the lattice momenta implied by our discretization
\begin{equation}
  \hat{p}_{\mu} = \frac{2}{a} \sin \left( a p_{\mu} / 2 \right)
\end{equation}
have entered.
In this formula $v$ and $m$ are the renormalized vacuum expectation value
and the renormalized mass and $Z$ is a multiplicative renormalization factor.
We prefer to define $v$ from a zero momentum contribution to the unsubtracted
two point function rather than from the direct expectation value $\langle \varphi \rangle$ 
because
we thus avoid subleties with an otherwise necessary symmetry breaking external field and
we prepare the extension of the scheme to a finite size system.

On a finite torus -- lattice or continuum -- 
all momentum components get quantized to integer multiples of $2\pi/L$.
We now focus on three momenta with the smallest
mutually differing values of $p^2$
\begin{eqnarray}
p&=&(0,0,0,0)\\
p_{\ast}&=&(1,0,0,0)\frac{2\pi}{L}\\
p_{\ast\ast}&=&(1,1,0,0)\frac{2\pi}{L},
\end{eqnarray}
for which we enforce (\ref{RNsmallp}) as exact equality
and then solve for $Z,m,v$.
The result is
\begin{equation}
  z^2 = \left( m L \right)^2 = \frac{G \left( p_{\ast \ast} \right)
  \hat{p}^2_{\ast \ast} L^2 - G \left( p_{\ast} \right) \hat{p}^2_{\ast}
  L^2}{G \left( p_{\ast} \right) - G \left( p_{\ast \ast} \right)}
\end{equation}
and
\begin{equation}
  w^2 = \left( v L \right)^2 = \frac{G \left( 0 \right)}{G \left( p_{\ast}
  \right)}  \frac{1}{L^2 \hat{p}^2_{\ast} + z^2} - z^{- 2} .
\end{equation}
where we have introduced the dimensionless finite size scaling
quantities $z=mL$ and $w=vL$. In addition we have
substitued
\begin{equation}
\delta^4(p) \to L^4 \delta_{p,0}
\end{equation}
for the finite size system.
A renormalized coupling in the broken phase is conveniently defined by the ratio
of mass to expectation value,
\begin{equation}
  g = \frac{3 m^2}{v^2} = \frac{3 z^2}{w^2}. \label{gdef}
\end{equation}
In an expansion around one of the degenerate minima
$g$ is seen to coincide with the bare coupling up to loop corrections.

In our numerical investigation we have restricted ourselves to the
limit $\lambda \to \infty$. Then the path integral over lattice fields $\varphi$
with weight $\exp(-S)$ reduces to the Ising model where we sum over cofigurations $\{\varphi(x)=\pm 1\}$.

The required observable $G(p)$ can be very conveniently estimated in the
loop representation \cite{Wolff:2008km} of the Ising model 
which is efficiently sampled by the worm algorithm \cite{prokofev2001wacci}.
In this ensemble the $\tanh(2\kappa)$ expansion of
\begin{equation}\label{calZ}
{\cal Z}=\sum_{u,v} \sum_{\varphi} \mathe^{2 \kappa \sum_{x,
  \mu} \varphi \left( x \right) \varphi \left( x + \hat{\mu} \right)} \varphi
  \left( u \right) \varphi \left( v \right)
\end{equation}
is sampled. As a consequence the distribution of $u$ and $v$
is related to the two point correlation,
\begin{equation}
  \langle \varphi \left( x \right) \varphi \left( 0 \right) \rangle =
  \frac{\langle \langle \delta_{x, u - v} \rangle \rangle}{\langle \langle
  \delta_{u, v} \rangle \rangle}
\end{equation}
where double angles refer to the average defined by 
(\ref{calZ}). Finally the desired Fourier transforms are given by
\begin{equation}
  G \left( p \right) = \frac{\langle \langle \mathe^{- i p \left( u - v
  \right)} \rangle \rangle}{\langle \langle \delta_{u, v} \rangle \rangle} =
  \frac{\left\langle \left\langle \prod_{\mu} \cos \left( p_{\mu} \left( u - v
  \right)_{\mu} \right) \right\rangle \right\rangle}{\langle \langle
  \delta_{u, v} \rangle \rangle}
\end{equation}
where the reflection invariance in each direction is used to get a real
representation in terms of cosines only.

\section{New data}
In comparison to \cite{Siefert:2014ela} we have substantially extended our simulations.
In table \ref{datatab} we compile our complete dataset. 
\begin{table}[!ht]
\centering
\input{tk/tab1.tex}
\caption{Simulation results for $z$ and $g$ at growing $L/a$ and
  for $\tilde{g}$ corrected to refer to $z^2 = 10$.\label{datatab}}
  \end{table}
Apart from some memory optimizations, our implementation of the worm algorithm is
a standard one. The only 4D field that we keep in memory, is the link field that
represents a graph of the $\tanh(2\kappa)$ expansion. Since it can only assume two
values per link, $V/2$ bytes suffice for its storage. Thus even our largest lattices 
fit comfortably into the memory of a standard desktop PC. 
It took roughly 43k core hours to generate our most expensive  ($L/a=160$) 
ensemble. This corresponds to 
$1.2\times 10^7\times V$ worm updates.

The coupling $\tilde{g}$ is related to the renormalized
coupling $g$ by
\begin{equation}
 \tilde{g} = g \frac1{(1-a^2 m^2/16)^2},
\end{equation}
i.~e. it only differs by a small lattice artefact.
The rationale \cite{Siefert:2014ela} is that the Callan Symanzik $\beta$ function for $g$
has a tree level artefact contribution which is absent for $\tilde{g}$.
For our $z^2=10$ data this amounts to a relative $1.25 L^{-2}$ correction that is
completely insignificant except for the smallest lattices.

\begin{figure}[htb!]
\centering
  \resizebox{0.5\textwidth}{!}{\includegraphics{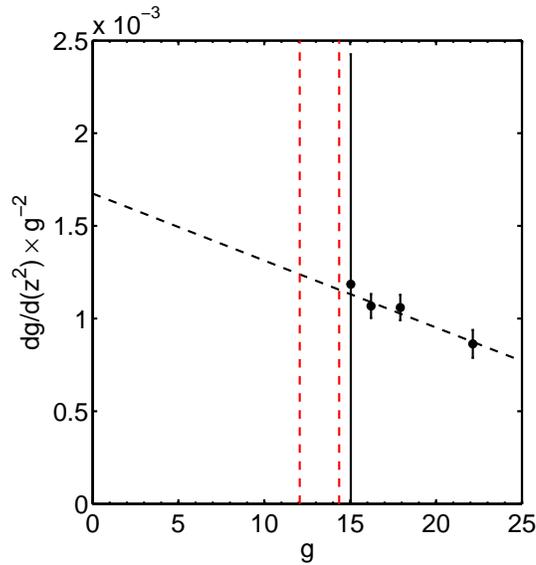}}
  \caption{Data and linear fit for $g^{-2} dg/dz$.\label{dgdz}}
\end{figure}
We had to tune $\kappa$ to approach $z^2=10$. Table \ref{datatab}
shows that we were often successful within our small statistical errors.
To implement the remaining tiny correction leading to the last column we
have numerically determined the derivative $d g/ d (z^2)$. 
This is relatively easy by using the relation
\begin{equation}\label{AKcon}
t \frac{d}{d t} \langle \langle A \rangle \rangle=
\langle \langle A K \rangle \rangle - 
\langle \langle A \rangle \rangle \langle \langle K \rangle \rangle
\end{equation}
which holds in the loop representation \cite{Wolff:2008km}.
In this formula $t\equiv \tanh(2\kappa)$ is used, $A$ can be
any $\kappa$-independent observable and $K$ is the total number
of links occupied by lines. We see in figure \ref{dgdz} that beyond
$L=64$ the connected correlation (\ref{AKcon}) is too noisy to get
a signal and we had to extrapolate with the shown linear fit. Its form is suggested
by perturbation theory.
We emphasize
however that any error in this procedure only affects a systematic
correction in the final results that itself is only of the order of the statistical error.

\begin{figure}[htb!]
\centering
  \resizebox{0.8\textwidth}{!}{\includegraphics{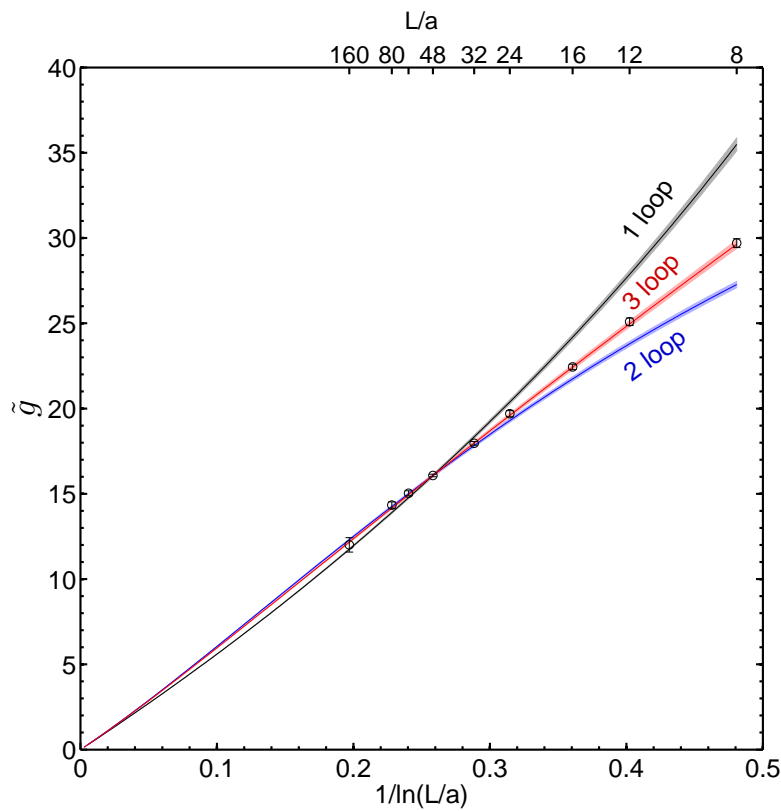}}
  \caption{Cutoff dependence of the coupling $\tilde{g}$.\label{gvslnL}}
\end{figure}
Our main result is now represented by figure \ref{gvslnL}. The curves
show the evolution with the perturbative renormalization group at 1,2 and 3 loop\footnote{
We have to note that the three loop term is taken from \cite{Luscher:1987ek}
and refers to $z=\infty$. According to experiences in the symmetric phase \cite{Weisz:2010xx}
the value for $z^2=10$ is expected to be very similar.}
precision
where the evolution is started (in both directions) from our most precise data point at $L/a=48$.
The shaded bands represent the small error of the initial value.
We see a perfect match of {\em all our data points} with the three loop evolution and thus
complete consistency with the (logarithmic) triviality scenario. 

\noindent Acknowledgements: We thank Peter Weisz
for discussions and the
Deutsche Forschungsgemeinschaft (DFG)
for support
in the framework of SFB Transregio~9.

\end{document}

%% file: tk/tab1.tex
\begin{tabular}{r D{.}{.}{7} D{.}{.}{6} D{.}{.}{6} D{.}{.}{6}}
\toprule
 \multicolumn{1}{c}{$L/a$} & \multicolumn{1}{c}{2$\kappa$}  &  
 \multicolumn{1}{c}{$z^2$} & \multicolumn{1}{c}{$g$} & \multicolumn{1}{c}{$
\tilde{g} |_{z^2=10}$}  \\
\midrule
    8  &  0.1524600  &  10.024(96)  &  29.13(30)  &  29.70(26)  \\
   12  &  0.1509920  &  10.008(98)  &  24.88(26)  &  25.09(22)  \\
   16  &  0.1504500  &  9.928(80)  &  22.30(19)  &  22.44(16)  \\
   24  &  0.1500460  &  9.974(98)  &  19.65(21)  &  19.70(18)  \\
   32  &  0.1498990  &  9.970(65)  &  17.93(13)  &  17.96(11)  \\
   48  &  0.1497900  &  10.189(53)  &  16.121(93)  &  16.074(77)  \\
   64  &  0.1497484  &  10.03(11)  &  15.03(18)  &  15.03(15)  \\
   80  &  0.1497294  &  10.10(14)  &  14.36(22)  &  14.34(19)  \\
  160  &  0.1497035  &  10.30(38)  &  12.06(49)  &  12.01(42)  \\
\bottomrule
\end{tabular}